\documentclass[useAMS,usenatbib]{mn2e}
\usepackage{graphicx,rotate,url,mathptmx,lscape,times, color,subfigure}
\def\gtsim{\mathrel{\hbox{\rlap{\hbox{\lower4pt\hbox{$\sim$}}}\hbox{$>$}}}}
\def\lesssim{\mathrel{\hbox{\rlap{\hbox{\lower4pt\hbox{$\sim$}}}\hbox{$<$}}}}
\def\Msunpyr{M$_{\odot}\,$yr$^{-1}$}
\def\Msun{M$_{\odot}$}
\def\erg{{\rm\thinspace erg}}

\def\Hz{{\rm\thinspace Hz}}

\def\km{{\rm\thinspace km}}

\def\Mpc{{\rm\thinspace Mpc}}
\def\Msun{\hbox{$\rm\thinspace M_{\odot}$}}

\def\Mpc{{\rm\thinspace Mpc}}
\def\s{{\rm\thinspace s}}
\def\ps{{\rm\thinspace s^{-1}}}

\def\yr{{\rm\thinspace yr}}

\def\simless{\mathbin{\lower 3pt\hbox
	{$\,\rlap{\raise 5pt\hbox{$\char'074$}}\mathchar"7218\,$}}} 
\def\simgreat{\mathbin{\lower 3pt\hbox
	{$\,\rlap{\raise 5pt\hbox{$\char'076$}}\mathchar"7218\,$}}} 

\def\ergps{\hbox{$\erg\s^{-1}\,$}}
\def\ergpsphz{\hbox{$\ergps\Hz^{-1}\,$}}

\def\kmps{\hbox{$\km\ps\,$}}
\def\kmpspMpc{\hbox{$\kmps\Mpc^{-1}\,$}}

\def\MsunpMpcc{\hbox{$\Msun\Mpc^{-3}\,$}}

\def\MsunpyrpMpcc{\hbox{$\Msun\yr^{-1}\Mpc^{-3}$}}

\def\Msunpyr{\hbox{$\Msun\yr^{-1}$}}

\def\pyr{\hbox{$\yr^{-1}\,$}}

\def\h0{\hbox{{\rm H}$^0$}}
\DeclareMathAlphabet{\vib}{OML}{cmm}{m}{it}

\def\h{$H_{\rm 160}$}

\def\clustername{{Cl 0218.3$-$0510}}
\begin{document}
\title[The impact of protocluster environments]{The impact of protocluster environments at $z = 1.6$}
\author[N.\,A.\,Hatch et al.]
       {\parbox[]{6.0in}
       {N.\,A.\,Hatch$^{1}$\thanks{E-mail: nina.hatch@nottingham.ac.uk},~E. A. Cooke$^{1}$,~S.\,I.\,Muldrew$^{2}$,~W.\,G.\,Hartley$^{3,4}$,~O.\,Almaini$^{1}$, C.\,J.\,Conselice$^{1}$, C.\,J.\,Simpson$^{5}$\\        \footnotesize
        $^1$School of Physics and Astronomy, University of Nottingham, University Park, Nottingham NG7 2RD\\ 
        	$^2$Department of Physics and Astronomy, University of Leicester, University Road, Leicester, LE1 7RH\\
	$^3$ETH Z\"urich, Institut f\"ur Astronomie, HIT J 11.3, Wolfgang-Pauli-Str. 27, 8093, Z\"urich\\
	$^4$Department of Physics and Astronomy, University College London, London, NW1 2PS\\
	$^5$Gemini Observatory, Northern Operations Center, 670 N.~A`oh\={o}k\={u} Place, Hilo HI 96720, USA
    }}
 \date{}
\pubyear{}
\maketitle

\label{firstpage}
\begin{abstract} 
We investigate the effects of dense environments on galaxy evolution by examining how the properties of galaxies in the $z=1.6$ protocluster \clustername\ depend on their location. We determine galaxy properties using spectral energy distribution fitting to 14-band photometry, including data at three wavelengths that tightly bracket the Balmer and 4000\AA\ breaks of the protocluster galaxies. We find that two-thirds of the protocluster galaxies, which lie between several compact groups, are indistinguishable from field galaxies. The other third, which reside within the groups, differ significantly from the intergroup galaxies in both colour and specific star formation rate. We find that the fraction of red galaxies within the massive protocluster groups is twice that of the intergroup region. These excess red galaxies are due to enhanced fractions of both passive galaxies (1.7 times that of the intergroup region) and dusty star-forming galaxies (3 times that of the intergroup region). We infer that some protocluster galaxies are processed in the groups before the cluster collapses. These processes act to suppress star formation and change the mode of star formation from unobscured to obscured.

\end{abstract}


\section{Introduction}

In the local Unvierse there are clear correlations between the environment of a galaxy and its star formation rate (SFR), morphology and colour, such that dense environments tend to host galaxies with suppressed SFRs, red colours and early-type morphologies \citep[e.g.][]{Dressler1980, Balogh1998, Kauffmann2004}. The result is that local clusters and groups are graveyards of passively evolving galaxies, gradually fading as their stellar populations age. 

As we observe further out in the Universe, the SFR--density, colour--density and morphology--density relations become less clear \citep[e.g.][]{Cucciati2006,Cooper2008,Grutzbauch2011,Ziparo2013}. Beyond $z\sim1.4$, clusters host large numbers of highly star forming galaxies \citep[e.g.][]{Brodwin2013}, which suggest that environmental quenching of star formation does not play as strong a role at these high redshifts. However, these dense environments still host a larger fraction of passive and red galaxies, which indicates that galaxy properties still depend on their environment even at high redshift where environmental quenching is less influential \citep{Chuter2011,Quadri2012}.

In this paper we study the protocluster \clustername\ at $z=1.6233$ \citep{Papovich2010,Tran2015}.  This protocluster is one of the densest known regions of the early Universe. The forming cluster has an X-ray-derived main halo mass of $5.7\times10^{13}$\Msun\ \citep{Tanaka2010}, and it is likely to grow into a $3\times10^{14}$\Msun\ cluster by the present day \citep{Hatch2016}.  
The galaxy population of the main group of the \clustername\ protocluster has been extensively studied due to its location in the well observed SXDS-UDS field, and partial coverage by CANDELS {\it HST} observations. The nascent cluster has an enhanced density of star formation \citep{Tran2010}, but also an enhanced passive galaxy fraction \citep{Quadri2012} relative to the field. The member galaxies are likely to be undergoing accelerated galaxy growth through mergers, which is supported by evidence of an enhanced merger rate \citep{Lotz2013} and larger galaxy sizes \citep{Papovich2012} compared to the field.

In this paper we investigate not only the main group, but the whole central 10 co-moving Mpc of the protocluster. We improve on previous studies by using an inventory of both star forming and passive protocluster galaxies, as well as detailed maps of the local density within the protocluster, to understand how galaxy properties correlate with environment before the cluster has fully collapsed. 
In Section \ref{method} we describe the method for locating the protocluster galaxies and determining their properties. In Section \ref{results} we compare the properties of the control field galaxies to the protocluster galaxies, and investigate the properties of galaxies as a function of their location within the protocluster. We discuss which protocluster environments affect galaxy properties and describe these environmental effects in Section \ref{discussion}, with conclusions following in Section \ref{conclusions}.

We use AB magnitudes throughout and a $\Lambda$CDM flat cosmology with $\Omega_M=0.315$, $\Omega_\Lambda=0.685$ and $H_0=67.3$ \kmpspMpc\ \citep{Planckcosmology2014}. 

\section{Method}
\label{method}
\subsection{Protocluster galaxy catalogue}
\label{catalogue}
We select a sample of 143 candidate protocluster galaxies and 88 control field galaxies from a  $7.5\arcmin\times7.5\arcmin$  field of view around \clustername\ using photometric redshifts. The center of the field of view lies at 02h18m28.75s, -05d09m52.19s. Full details of the data and methods used to select the protocluster galaxies are provided in \citet{Hatch2016} which we summarise here.

A $K$--selected $UBVRi^{\prime}z^{\prime}JHK$[3.6][4.5] catalogue of the galaxies around \clustername\ was created by \citet{Hartley2013} as part of the eighth data release of the near-infrared UKIDSS Ultra Deep Survey (UDS). We supplemented this photometry  with additional narrow and medium-band images at $0.9527\micron$, 1.02\micron\ and 1.06\micron, and broad-band $J$ and $Ks$ images from FORS and HAWK-I at the VLT. The data at 0.9-1.1$\micron$ sample the Balmer and 4000\AA\ breaks of the protocluster galaxies at multiple points, allowing us to derive accurate photometric redshifts.

Photometric redshifts were determined for each of the 3019 galaxies in the field of view using the {\sc eazy} code \citep{Brammer2008}.  Due to the tight sampling of the Balmer and 4000\AA\ breaks we obtained a ($z_{phot}-z_{spec}$) dispersion of $\Delta z/(1+z)=0.013$ for galaxies at $z\sim1.6$.

Protocluster galaxies were chosen as those that had more than a 50\% probability of lying within $z_{pc}\pm0.068$ and $>90$\% probability of lying between $z_{pc}\pm0.17$, where $z_{pc}=1.6233$ \citep{Tran2015}.  These 143 protocluster galaxies are the \lq Goldilocks\rq\ sample described in \citet{Hatch2016}, which is a compromise between choosing a clean and complete protocluster galaxy sample. By comparison to the spectroscopic sample of \citet{Tran2015} we estimate that the Goldilocks protocluster sample has only 12 per cent contamination of field galaxies \citep{Hatch2016}. A control field galaxy sample was selected from the same field of view and with the same criteria, but at redshifts $z=1.45$ and 1.81; the control field galaxy sample was selected from a volume of approximately twice the size of the protocluster field. The low photometric redshift uncertainties mean that these three redshift bins are well separated.

\begin{figure}
\includegraphics[height=1.1\columnwidth, angle=-90]{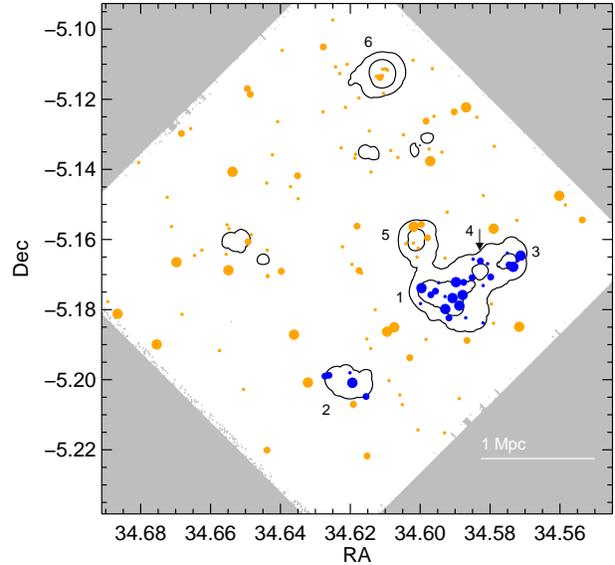}
\caption{\label{fig:map_groups} The distribution of massive group (blue circles) and intergroup (orange circles) galaxies within the protocluster at $z=1.6233$. The contours mark regions with high galaxy density and delimit the 6 galaxy groups of the protocluster, which are labelled. Groups 5 and 6 have stellar masses $<10^{11}$\Msun\ and we do not classify them as massive groups. The size of the circles are proportional to the stellar mass of the galaxy with larger symbols indicating more massive galaxies.}
\end{figure}

The \clustername\ protocluster consists of six galaxy groups surrounded by a dense sea of intergroup galaxies (see Fig.\,\ref{fig:map_groups}). In \citet{Hatch2016} we showed that the protocluster consisted of two structural differences in comparison to the control field.  First, the density of the intergroup galaxies was $2 - 2.5$ times as dense as the control field. Secondly, four of the six groups were much more massive than groups observed in the control field, each containing total observed stellar masses from $10^{11}$\Msun\ to $10^{11.93}$\Msun. Details of the location, number of galaxies and total stellar mass content of the six groups in the protocluster are provided in table 2 of \citet{Hatch2016}.

In this work we examine the properties of galaxies in the four most massive protocluster groups and those in the intergroup region separately to determine the effect these different structures have on the properties of the protocluster galaxies. For this analysis we combine the galaxies which belong to the two lowest mass protocluster groups with the intergroup galaxies (see Fig.\,\ref{fig:map_groups}). This is because the total stellar mass of these groups is $<10^{11}$\Msun, which is less than the stellar mass of single galaxies within the intergroup region, so the total stellar mass density of these low mass groups is comparable to the intergroup region. Furthermore, the control field contains similar mass groups so these low-mass groups are not unique signatures of the protocluster. Hereafter we refer to the low-mass group and intergroup galaxies as the intergroup galaxies for simplicity.

Our observations of the protocluster trace an area of $10.2\times10.2$\,co-moving Mpc and an approximate depth of 34\,co-moving Mpc \citep{Hatch2016}. The intergroup region will extend across almost all of that volume, but the four most massive groups (labelled 1-4 in Fig.\,\ref{fig:map_groups}) cover only 5.7\% of the $10.2\times10.2$\,co-moving Mpc area of the protocluster. If the groups have the same depth as width then they only fill 0.4\% of the observed protocluster volume. There are 110 galaxies in the intergroup region, and 33 galaxies in the four most massive groups. Hence, the average volume density of galaxies is 75 times greater in the four most massive groups than between the groups.

It is highly likely that all of the groups will collapse into the cluster, but the probability of the intergroup galaxies becoming cluster members diminishes with radius, so galaxies on the eastern edge of the field of view have less than 60\% chance of becoming cluster galaxies \citep{Hatch2016}. Simulations of similar mass protoclusters suggest that approximately 37 of the 143 protocluster galaxies are not likely to become cluster members;  however, it is impossible to determine exactly which are outliers. We therefore treat all 143  galaxies as potential protocluster members when comparing the properties of the protocluster galaxies to control field galaxies. This should not greatly affect our results as \citet{Contini2016} showed that there is no significant difference in the properties of the protocluster galaxies and the outliers that lay within the protocluster volume. Furthermore, we checked whether the properties of intergroup galaxies varied as a function of distance from the most massive group as the outer region is likely to contain a higher fraction of outliers, but we found no difference in the stellar masses, SFRs or colours. 

\subsection{Measuring galaxy properties}

We derived stellar masses and dust extinction ($A_V$) by fitting the $U$ to 4.5\micron\ photometry with stellar population models using the {\sc fast} code \citep{Kriek2009a}. The redshifts of the protocluster members were fixed to $z=1.6233$ whilst the redshifts of the field galaxies were fixed to redshifts (m$_2$) produced by {\sc eazy}. The dust attenuation ($A_V$) was allowed to range between 0 and 4\,mag. Further details of the fitting procedure are provided in \citet{Hatch2016}. 

Dust-corrected SFRs were derived using the \citet{Kennicutt1998} conversion from 2800\AA\ (assuming a Chabrier initial mass function): ${\rm SFR (}\Msun yr^{-1}{\rm )} = 8.24 \times 10^{-29} L_{2800} {\rm(}\ergpsphz \rm{)}$, after the observed UV luminosity was corrected for dust extinction. We estimate the amount of dust attenuation at 2800\AA\ ($A_{2800}$) using the $A_V$ derived by {\sc fast} and converting it to $A_{2800}$ using the \citet{Kriek2013} average dust law (with $E_b= 1$ and $\delta=0.1$). We compared the dust-corrected UV-based SFRs to SFRs derived from MIPS $24\mu $m fluxes (using \citealt{Rujopakarn2013} corrections) from the {\it Spitzer} UKIDSS Ultra Deep Survey (PI Dunlop). We corrected the UV-based SFRs to be the MIPS SFRs if the MIPS SFRs were $>60$\Msunpyr\ and deviated by more than a factor of two from the UV-based SFRs. Generally the SFRs derived from MIPS data agreed very well with the dust-corrected UV SFRs, and we only updated 9 protocluster SFRs and 3 control field SFRs.

We define passive galaxies as those which have specific star formation rates (sSFRs) less than 1\,Gyr$^{-1}$,  i.e., $\log~{\rm sSFR}/\pyr <-9$.  The results of this work are insensitive to the exact definition of passive galaxies since we are comparing galaxies within the same dataset and with the same definition of passivity.

\subsection{Catalogue Completeness} 
\label{completeness}

There are two completeness issues that we consider for these data. First, there is the flux completeness of the $K-$selected catalogue. The catalogue is $>99$\% complete at $K=24.3$\,mag \citep{Hartley2013}.  This is equivalent to $10^{9.7}$\Msun\ for a passive, red galaxy at $z=1.62$. The second issue is the level of completeness due to the definition of protocluster membership. The completeness of protocluster membership depends on both magnitude and galaxy colour, with fainter and redder galaxies being less likely to be selected as protocluster galaxies than brighter and bluer galaxies. This is because the protocluster members are selected based on their redshift probability distribution functions, $P(z)$, and fainter and redder galaxies have broader $P(z)$ distributions due to higher fractional flux errors.

To measure how the protocluster membership completeness varies with flux and colour we scaled the best fitting spectral energy distributions of all 143 galaxies in the protocluster sample such that the $K$ magnitudes varied from 20.0 to 24.5\,mag. Uncertainties were added to the flux in each band by assuming Gaussian errors and shifting the simulated galaxy fluxes within their error bounds.  The uncertainties on the fluxes were kept at the original level.  The photometry was run through the {\sc eazy} photometric redshift code to obtain the full redshift probability distributions of each simulated galaxy, and then run through the SED-fitting code {\sc fast} to determine stellar masses. 
 
The completeness of the protocluster catalogue as a function of both stellar mass and colour was determined as the fraction of simulated galaxies within a certain mass and colour range that obeyed the protocluster's redshift probability criteria ($>50$\% probability of lying within $z_{pc}\pm0.068$ and $>90$\% probability of lying between $z_{pc}\pm0.17$). The 90\%, 75\%, 50\% and 25\% completeness curves are plotted on Figs\,\ref{fig:CMassD_field}-\ref{fig:red_fraction} and \ref{fig:dusty_fraction}. The $>99$\% $K-$band flux completeness limit (24.3 mag) is also shown on these figures as the solid black line. There is no reason to believe that the incompleteness would be systematically in favour or against a particular galaxy type in different environments, so we assume the level of protocluster member incompleteness is the same in group and intergroup regions.

\section{Results}
\label{results}

\subsection{Comparison of intergroup and control field galaxies}

\label{fil_field_compare}
\begin{figure}
\includegraphics[height=1\columnwidth, angle=-90]{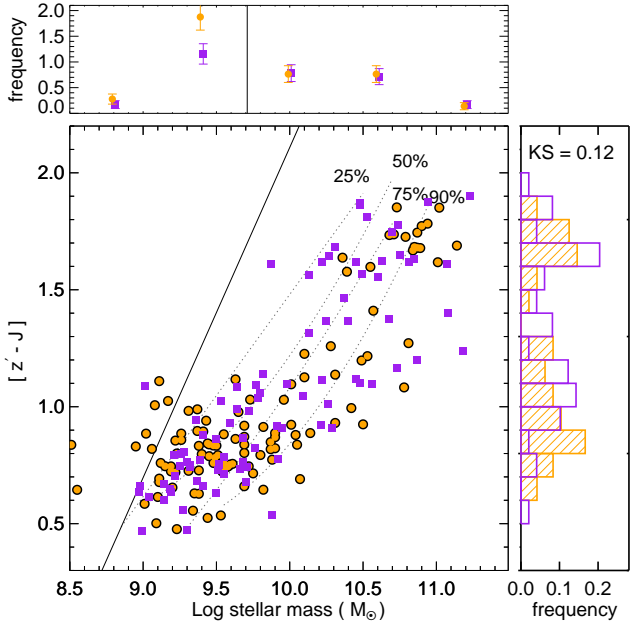}
\caption{\label{fig:CMassD_field} A comparison of the $z^{\prime}-J$ colours and stellar masses of control field galaxies (purple squares) and intergroup protocluster galaxies (orange circles). The solid line in the main panel is the limit dictated by the 99\% flux completeness of the $K-$selected catalogue. The dotted lines are the completeness limits caused by the stringent photometric redshift criterion for protocluster and control field membership (see Section~\ref{completeness} for details). The colours of the field galaxies have been K-corrected to $z=1.62$. The distributions of stellar masses (upper panel) have been normalised such that both populations have the same frequency of galaxies above the detection limit of $M>10^{9.7}$\Msun (marked by the solid line), whilst the right-hand panel only compares the colour distributions of galaxies above the detection limit.
 }
\end{figure}

In this section we compare 92 control field galaxies to the 110 intergroup galaxies in the protocluster (marked as orange circles in Fig.\,\ref{fig:map_groups}).  Fig.\,\ref{fig:CMassD_field} compares the colours and stellar masses of intergroup galaxies (orange circles) and control field galaxies (purple squares). The  $z^{\prime}-J$ colour spans across the Balmer and 4000\AA\ break so this colour separates passive and dusty galaxies from unobscured star-forming galaxies.  The $z^{\prime}-J$ colours of the control field galaxies have been K-corrected to $z=1.62$. 

The distributions of control field and intergroup galaxies in colour-mass space are very similar. Galaxies in both samples populate a red sequence at $z^{\prime}-J\sim1.7$ and a blue cloud at $0.5<z^{\prime}-J<1.2$. The sample of control and intergroup galaxies are incomplete at almost all masses due to the assertion that the galaxies must have a high probability of lying within the control field redshift range or the protocluster. Dashed lines marking where the protocluster galaxies are 90, 75, 50 and 25\% complete are shown on Fig.\,\ref{fig:CMassD_field}. The control field galaxies will have slightly different completenesses, shifted to brighter magnitudes if they lie at $z=1.81$, and fainter if they lie at $z=1.45$.  

Kolmogorov-Smirnov (KS) tests comparing the stellar mass distributions and colour distributions result in probabilities of 0.38 and 0.06, respectively, that the samples are drawn from the same underlying distribution. Due to the 99\% flux completeness limit we are only able to detect red galaxies with stellar masses $>10^{9.7}$\Msun. Taking only galaxies above this mass limit results in even higher KS probabilities (0.77 and 0.12, respectively). Furthermore, the red fraction of galaxies with $M>10^{9.7}$\Msun\ is comparable in the two regions, at 38$\pm10$\% of galaxies in the protocluster intergroup region and 44$\pm12$\% in the control field. This means there are no statistically significant differences between the properties of the intergroup and control field galaxies. We conclude that there is no evidence of the protocluster intergroup environment influencing the masses or colours of the intergroup galaxies. Intergroup protocluster galaxies appear to be pristine infall galaxies with the same properties as \lq field\rq\ galaxies. 

This comparison should be treated with caution since the control field galaxies lie at a slightly different redshift to the protocluster galaxies, with the vast majority of them at $z=1.45$. The completeness of the control field galaxies is likely to be slightly less for a given mass than the protocluster galaxies due to the larger uncertainties on their photometric redshifts (see \citealt{Hatch2016} for further details). This effect shifts the dashed completeness lines in Fig.\,\ref{fig:CMassD_field} slightly to the right. Countering this, the difference in redshift results in a 0.3\,mag luminosity difference, hence lower-mass and passive galaxies are more easily identified in the control field than in the protocluster. This second effect shifts the dashed completeness lines back to the left, so the two effects correct for each other. Overall, these differences in the selection of protocluster and control field galaxies means we should be cautious when comparing these two populations.

\subsection{Comparison of the intergroup and group galaxies within the protocluster}
\label{fil_group_compare}

\begin{figure}
\includegraphics[height=1\columnwidth, angle=-90]{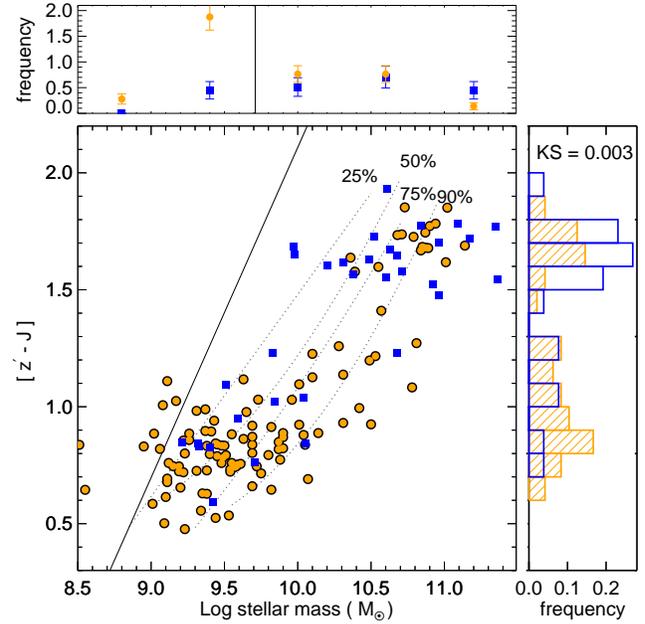}
\caption{\label{fig:CMassD} A comparison of the $z^{\prime}-J$ colours and stellar masses of the 33 galaxies within the four most massive groups (blue squares) and the 110 intergroup protocluster galaxies (orange circles). The solid and dotted lines in the main panel are described in Fig.\,\ref{fig:CMassD_field}.  The right-hand panel compares the colour distributions of galaxies above the detection limit, i.e., $M>10^{9.7}$\Msun, and shows that there is a significantly higher fraction of red galaxies in the groups. The top panel compares the stellar mass distributions (normalised to match the frequency of galaxies above the detection limit, which is marked by the solid line) and shows that intergroup region has a significant excess of low-mass ($M<10^{9.5}$\Msun) blue galaxies in comparison to the group galaxies. 
}
\end{figure}

We now explore the influence of the protocluster group environment on galaxy properties by comparing the properties of the 110 intergroup protocluster galaxies to the 33 galaxies within the four groups with total observed stellar masses $>10^{11}$\Msun. In Section \ref{fil_field_compare} we showed that the galaxies between the protocluster groups have the same properties as those in the control field. It is preferable to compare the protocluster group galaxies to the intergroup galaxies rather than the control field galaxies since the intergroup galaxies lie at exactly the same redshift, and so the selection biases and mass completeness limits are identical for these samples.

\subsubsection{Comparison of colours and stellar masses}

We compare the group and intergroup protocluster galaxies in Fig.\,\ref{fig:CMassD}. The dotted lines mark the 90, 75, 50 and 25\% completeness limits of the data. Although the data are incomplete at all masses, they are incomplete in exactly the same way for the two populations as both the group and intergroup galaxies are at the same redshift and have been selected in the same way. Therefore we can directly compare the distribution of galaxies in this colour-mass space for these two populations, keeping in mind the regions of colour-mass space that are sparse due to incompleteness.

Both group and intergroup galaxies populate a blue cloud and red sequence, but the relative distribution of galaxies within these two regions differs for the two environments. Considering the full galaxy population shown in Fig.\,\ref{fig:CMassD} the KS probabilities comparing the mass and colour distributions are 0.001 and $7\times10^{-6}$, respectively. These low probabilities mean that the galaxies in the massive groups have significantly different properties to those that reside in the less dense regions of the protocluster or in the field. When we consider only those galaxies with stellar masses above the $10^{9.7}$\Msun\ limit we detect no difference in the stellar mass distribution (KS~probability\,=\,0.43), but we still find significant differences in colour (KS~probability\,=\,0.003; see the right-hand panel of Fig.\,\ref{fig:CMassD}). Thus we find that the group galaxies are redder than the intergroup galaxies. It is unclear from these data whether the low-mass galaxies in the groups are not visible because they are redder than those in the intergroup region, and therefore fall below the flux detection limit, or whether they are simply not present, having merged with larger galaxies in the denser environment. We discuss these low-mass galaxies in more detail in Section\,\ref{faint_blueies}.

\begin{figure}
\includegraphics[height=1\columnwidth, angle=-90]{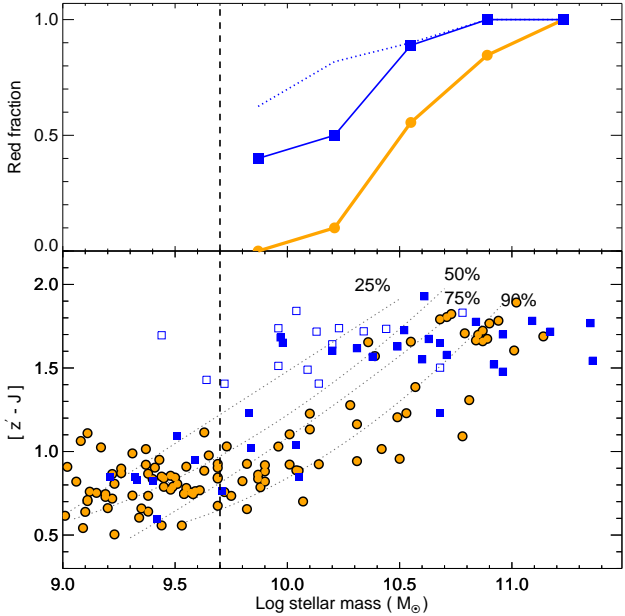}
\caption{\label{fig:red_fraction}The top panel shows the red fraction of massive group (blue) and intergroup (orange) protocluster galaxies. The bottom panel shows the colour-mass relation (symbols defined in Fig.\,\ref{fig:CMassD}). The open blue squares are red galaxies  that lie within the area defined as massive groups but do not match the protocluster redshift criteria, and are therefore not classified as protocluster members. The blue dotted line in the top panel shows the red fraction including these galaxies, and is therefore an upper limit, whilst the solid lines are lower limits (see text for details). The black vertical dashed line marks the mass limit below which we cannot detect red protocluster galaxies. The red fraction decreases with decreasing stellar mass in both environments, but it decreases more rapidly in the intergroup environment.}
\end{figure}

\begin{figure}
\includegraphics[height=1\columnwidth, angle=-90]{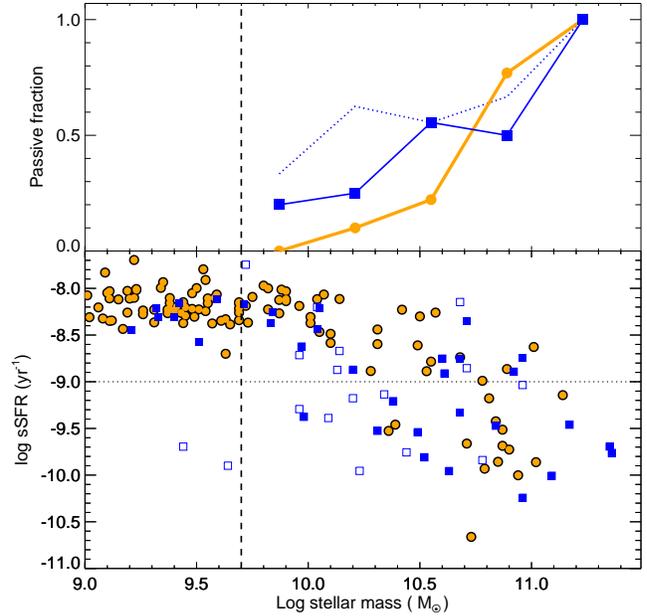}
\caption{\label{fig:passive_fraction}The top panel shows the passive fraction of massive group (blue) and intergroup (orange) protocluster galaxies. The bottom panel shows the position of galaxies in the sSFR-mass plane. Symbols and lines are  defined in Fig.\,\ref{fig:red_fraction}. The horizontal dotted black line marks the dividing line between passive and star forming galaxies. The passive fraction decreases with decreasing stellar mass, and the largest difference occurs at the lower-mass end, where the passive fraction of group galaxies is larger than that of the intergroup galaxies. }
\end{figure}

\begin{figure}
\includegraphics[height=1\columnwidth, angle=-90]{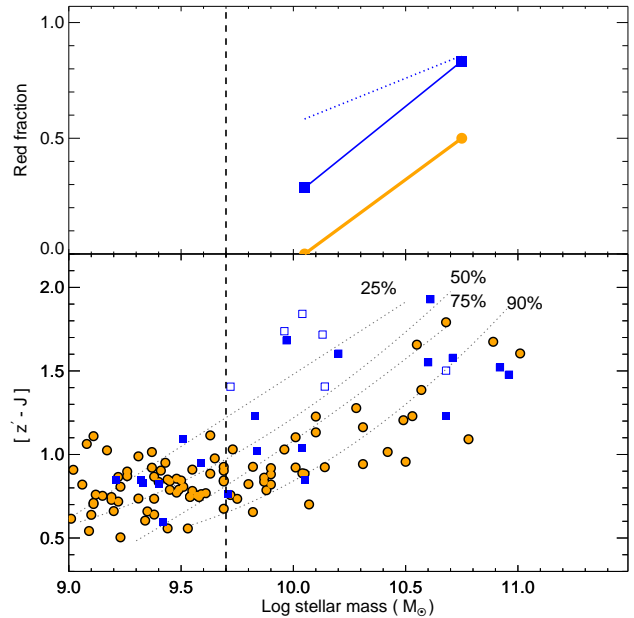}
\caption{\label{fig:dusty_fraction}The top panel shows the red fraction of star forming galaxies within massive groups (blue) and the intergroup region (orange). The bottom panel shows the colour-mass relation. Symbols and lines are the same as those defined in Fig.\,\ref{fig:red_fraction}. The red fraction decreases with decreasing stellar mass, and there is an excess of red star-forming galaxies in the massive groups at all stellar masses.}
\end{figure}

\subsubsection{Red and passive galaxy fractions}
In this section we investigate the cause of the excess red galaxies in the protocluster groups and we only consider galaxies with $\log{M/M_{\odot}}>9.7$, i.e. with masses greater than the K-band flux limit for a red galaxy.

The fraction of red galaxies, defined as galaxies with $z^\prime - J>1.4$, is $77\pm23\%$ in the groups, and $38\pm10\%$ in the intergroup region. Thus the groups contain twice the red fraction compared to intergroup galaxies. As described in Section\,\ref{completeness}, the completeness of our protocluster sample is a function of galaxy colour at all stellar masses. Since the completeness of the red galaxies is lower than the blue galaxies, the fractions we quote should be considered lower limits. However, it is safe to directly compare the red and passive fractions of the group and intergroup galaxies as these samples have exactly the same completeness limits.

In Fig.\,\ref{fig:red_fraction} we investigate the mass dependance of the red fraction. The solid points are as described in Fig.\,\ref{fig:CMassD}. The open blue squares are red galaxies  ($z^\prime - J >1.4$) that lie within the area defined as massive groups (see Fig.\,\ref{fig:map_groups}) but do not match the protocluster redshift criteria (see Section\,\ref{completeness}), and are therefore not defined as protocluster members. These galaxies may be interlopers or they may be misclassified protocluster members. We use these galaxies to define upper limits to the red and passive fractions (shown as dotted blue lines in the upper panels of Figs.\,\ref{fig:red_fraction}-\ref{fig:dusty_fraction}). Therefore the true red and passive fractions of the group galaxies will lie between the solid and the dotted lines of Figs.\,\ref{fig:red_fraction}-\ref{fig:dusty_fraction}.

We find that the red fractions increase with stellar mass in all environments. Whilst this may be due to the dependance of incompleteness on galaxy colour, we also find that the upper limit of the red fraction increases with stellar mass in the group environment. This result is in agreement with \citet{Rudnick2012}, who reported a lack of faint red galaxies in the central region of this protocluster. The red fraction is very similar (at almost 100\%) for the most massive group and intergroup galaxies. At lower masses the red fraction diverges with the intergroup fraction decreasing more rapidly with decreasing mass. So the gulf between the red fraction in different environments expands at lower masses.

\begin{figure}
\includegraphics[height=1.1\columnwidth, angle=-90]{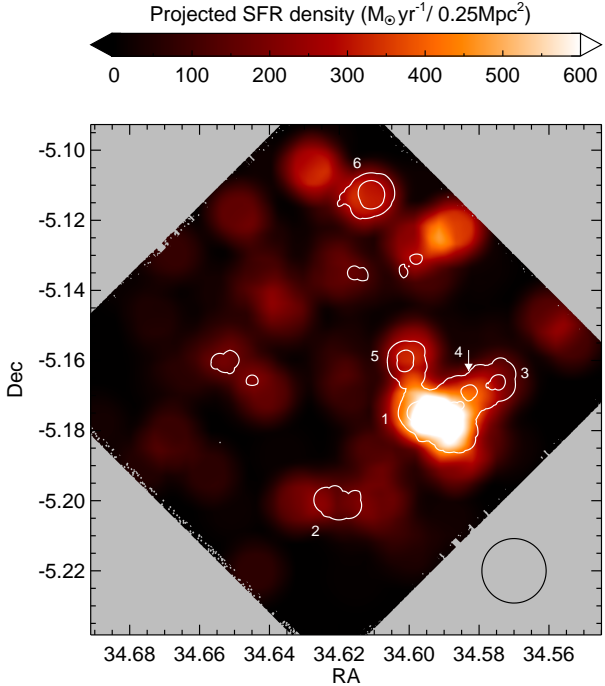}
\caption{\label{fig:sfr} A map of the projected SFR density in the protocluster. The colour scheme indicates the total SFR in \Msunpyr\ within a region of $0.25$\,Mpc$^{2}$ (shown by the black circle in the bottom right corner of the map). White contours mark the galaxy groups in the protocluster. The SFR is strongly enhanced in the most massive group of the protocluster (group 1)}
\end{figure}

To investigate the origin of the additional red galaxies within the groups we examine the passive galaxy fraction and distribution of sSFRs in Fig.\,\ref{fig:passive_fraction}. In total the groups have a passive fraction of $50\pm17\%$, compared to only $29\pm9\%$ in the intergroup region. The passive fraction increases with increasing stellar mass, similar to the red fraction. The upper limit of the passive fraction increases with stellar masses more steeply than the red fraction. Furthermore, the difference between the passive fraction in the two environments expands at lower masses, although with less significance than the red fraction. The sSFRs of the group galaxies differ from the intergroup galaxies with a KS probability of 0.02. 

The passive fraction is below the red fraction in almost all mass bins in both environments, which means that some of the red galaxies must be star-forming. In Fig.\,\ref{fig:dusty_fraction} we plot the red fraction of only the star forming galaxies ($\log {\rm sSFR}>-9$\pyr). There is a higher fraction of red star-forming galaxies in the groups than in the intergroup region at all masses. Whilst $54\pm 20\%$ of the star forming galaxy population in groups are red, only $15\pm 7\%$ of the star forming population are red in the intergroup region. In total,  $30\pm 12\%$ of the group galaxies are red star-forming galaxies compared to only $11\pm 5\%$ in the intergroup region.  Thus the groups contain an enhanced fraction of red star-forming galaxies. 

In summary, the groups have twice the red fraction of the intergroup region. This is due to an enhanced passive fraction (by a factor of 1.7) and a higher fraction of star forming galaxies exhibiting red colours (a factor of 3 enhancement). The excess of red and passive galaxies increases with decreasing galaxy mass, but the excess of red star-forming galaxies appears to be similar at all masses probed. The population that is depleted in the groups are blue star-forming galaxies.

\subsection{The distribution of star formation within the protocluster}

The \clustername\ protocluster is a highly star-forming structure. Combining the SFR of all protocluster galaxies with stellar masses above our mass completeness limit of $10^{9.7}$\Msun\ results in a total SFR of more than 4000\Msunpyr\ within the $10.2\times10.2$\,co-moving Mpc area covered by our images. This translates into a global star formation rate density ($\rho_{\rm SFR}$) of 1.1\MsunpyrpMpcc, which is ten times greater than the field density at this redshift \citep{Cucciati2012, Magnelli2013}. Thirty percent of the star formation occurs within the massive groups of the protocluster ($\sim1200$\Msunpyr), implying that the star formation rate density is $\rho_{\rm SFR}= 82$\MsunpyrpMpcc\ within the groups, which is over 800 times the field density. These values are summarised in Table \ref{tab:ssfr}. The protocluster is a highly star forming region of the Universe that produces an order of magnitude more stars per unit volume than an average part of the Universe.\footnote{If we include galaxies below the $10^{9.7}$\Msun\ detection limit the total SFR of the protocluster increases to 5300\Msunpyr\ of which only a quarter (1300\Msunpyr) occurs in the massive groups. The average  star formation rate density is $\rho_{\rm SFR}=1.5$\MsunpyrpMpcc, and the groups have $\rho_{\rm SFR}=91$\MsunpyrpMpcc.}

\begin{table}
\begin{tabular}{|l|c|c|c|l|}
\hline
&Groups &  Intergroup & Whole & Field  \\
&& & protocluster&   \\
 \hline
SFR/\Msunpyr & 1168& 2826	& 4023&  --	\\
 $\rho_{\rm SFR}$/\MsunpyrpMpcc &81.9	&0.8	&  1.1	& $0.11_{+0.02}^{+0.06}$~$^{R1}$	\\
 $\rho_{\rm*}$/\MsunpMpcc &107	&0.53	&  0.96 & $0.10\pm0.01$~$^{R2}$	\\
$\log<$sSFR$>$/\pyr &-9.11	&-8.78	& -8.90	& $-8.65$	~$^{R3}$\\
\hline

\end{tabular}
\caption{\label{tab:ssfr} Star formation and stellar mass properties of different regions in the protocluster.
Field values are taken from $^{R1}$~\citet{Cucciati2012}, $^{R2}$~\citet{Muzzin2013b}, and $^{R3}$~\citet{Tasca2015} who evaluated the sSFR for $>10^{10}$\Msun\ galaxies. The Groups category only includes the 4 most massive protocluster groups, whilst the intergroup category includes the two least massive groups with stellar masses $<10^{11}$\Msun.
}

\end{table}

To investigate the cause of this high $\rho_{\rm SFR}$ we examine the average stellar mass density and sSFR in different parts of the protocluster and compared them to an average field (see Table.\,\ref{tab:ssfr}). The average stellar mass density of the protocluster is ten times greater than the field, but the stellar mass is highly concentrated in the massive protocluster groups. This can be seen in the projected stellar mass density map presented in Fig.\,5 of \citet{Hatch2016}. The average stellar mass density of the four most massive groups is over a thousand times greater than the field density, whereas the intergroup region has a lower stellar mass density of only five times the field density (see Table\,\ref{tab:ssfr}).

This high mass concentration means that the average sSFR for the protocluster is slightly below the field rate at this redshift. The intergroup region has a comparable rate, whilst the average sSFR is lower in the groups than the field (see Table\,\ref{tab:ssfr}). Therefore the star formation rate per unit stellar mass is suppressed in the protocluster, with most of the suppression occurring in the massive groups. But because there is so much stellar mass in such a small volume, it still produces a very high SFR per unit volume.

Fig.\,\ref{fig:sfr} shows that star formation is centrally concentrated in the most massive group of the protocluster (group 1), confirming the result of \citet{Tran2010}. This map was created by summing the SFR of galaxies within 30\arcsec\ of each 0.25\arcsec\ pixel, and then smoothing the map with a boxcar average of 100 pixel width. The map is scaled to display the SFR within a 0.25\,Mpc$^{2}$ region. The total SFR within 1\,Mpc of the most massive protocluster galaxy is 1060\Msunpyr, which is in excellent agreement with \citet{Santos2014}, who measured the $\rho_{\rm SFR}$ using {\it Herschel} and {\it Spitzer}/MIPS infrared luminosities.

\begin{figure}
\includegraphics[width=1\columnwidth, angle=-90]{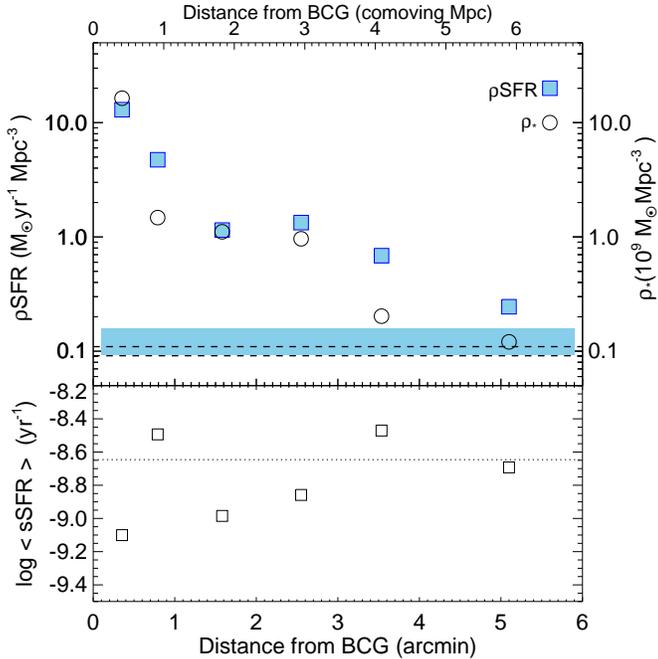}
\caption{\label{fig:radial_ssfr} The top panel shows the radial distribution of stellar mass density ($\rho_{*}$: open circles) and star formation density ($\rho_{\rm SFR}$: blue squares), whilst the bottom panel shows the radial distribution of the average sSFR within annuli of the protocluster. The horizontal dashed lines mark the field $\rho_{*}$ \citep{Muzzin2013b}. The shaded blue area marks the field $\rho_{\rm SFR}$  \citep{Cucciati2012}. The dotted horizontal line in the bottom panel marks the field sSFR for $>10^{10}$\Msun\ galaxies \citep{Tasca2015}. Both $\rho_{*}$ and $\rho_{\rm SFR}$ are greatly enhanced in the central regions of the protocluster, and reach the field level at $\sim$6\,co-moving Mpc from the central cluster galaxy. The sSFR is generally decreased relative to the field in the inner 3\,co-moving Mpc.}
\end{figure}

Groups are concentrated in the central region of the protocluster, so we explore how the presence of the groups affects the global properties of the protocluster through the radial distribution of the stellar mass density ($\rho_{*}$), $\rho_{\rm SFR}$ and sSFR in Fig.\,\ref{fig:radial_ssfr}. We take the origin as the most massive protocluster galaxy, which lies at RA=34.58976, DEC=-5.17217 (J2000). The depth of the field is taken to be 34\,co-moving Mpc \citep{Hatch2016}, which may reduce the observed central $\rho_{*}$ and $\rho_{\rm SFR}$ because the group galaxies are unlikely to be distributed over the entire 34\,co-moving Mpc depth of the protocluster.

We find that the central region of the protocluster has more than two orders of magnitude greater $\rho_{*}$ and $\rho_{\rm SFR}$ than the field \citep{Cucciati2012,Muzzin2013b}. Both $\rho_{*}$ and $\rho_{\rm SFR}$ decrease with radius and approach the field values at $\sim6$\,co-moving Mpc\, which in physical units is $\sim2.3$\,Mpc. The bottom panel of Fig.\,\ref{fig:radial_ssfr} shows the average sSFR ($\rho_{\rm SFR}/\rho_{*}$) within each annulus. Beyond 3\,co-moving Mpc the average sSFR is similar to the field value \citep{Tasca2015}. Within this radius the average sSFR is generally lower than the field, which indicates that star formation is suppressed within the protocluster. The exception is the 1\,co-moving Mpc annulus where the sSFR is slightly enhanced. This high value is due to a few highly star forming galaxies, including the most infrared luminous galaxy in the protocluster, which has a SFR of $\sim$250\Msunpyr\ (ID 16 in \citealt{Santos2014}).

In conclusion, we find that both $\rho_{*}$ and $\rho_{\rm SFR}$ are greatly enhanced in the protocluster, particularly within the groups where the $\rho_{\rm SFR}$ may be $\sim800$ times that of the general field. Both $\rho_{*}$ and $\rho_{\rm SFR}$ decrease with distance from the protocluster core, until they reach the field level at $\sim6$\,co-moving Mpc. The average sSFR is generally decreased in the central 3\,co-moving Mpc region compared to the outer regions, meaning that star formation is suppressed within the inner regions of the protocluster.

\section{Discussion}
\label{discussion}

\subsection{Which environment matters}
Protoclusters consist of galaxies that reside in dense groups, and galaxies that lie between these groups. These protocluster galaxies experience very difference environments, for example the stellar mass density in the intergroup region is only $0.5\times10^9$\Msun\Mpc$^{-3}$ compared to $107\times10^9$\Msun\Mpc$^{-3}$ in the groups.\footnote{We define groups as the structures that contain more than $10^{11}$\Msun\ stellar mass, so the low-mass groups 5 and 6 are included with the intergroup region.  The total stellar mass of these groups is less than some single galaxies in the intergroup region so their stellar mass density is likely comparable to the intergroup region.} Our results show that the intergroup galaxies do not differ from field galaxies, whereas the group galaxies differ in terms of their colours and star formation properties. This means that some of the galaxies that end up in clusters are preprocessed in groups within protoclusters before the cluster assembles.

Examining the protocluster as a single structure, we find that the sSFR decreases within the inner 3\,co-moving Mpc. This is because most of the massive groups are within this radius. These groups, which have an excess of passive galaxies, decrease the total sSFR within this radius. It is therefore not only the most massive protocluster group that matters, but rather all massive groups can affect the properties of their member galaxies, and contribute to pre-processing of the cluster population.

The global properties of the protocluster environment (defined as $\rho_{*}$ and $\rho_{\rm SFR}$) drop to the field level at a radius of 6\,co-moving Mpc. This is the radius beyond which there are few protocluster groups (only a single low mass group). Since the global environment is similar to the field at this radius, it is reassuring that we find the sSFR matches the field level also. Approximately 65\% of the galaxies at this radius will become cluster members by $z=0$ \citep{Hatch2016}. These protocluster galaxies, and those at larger radii, can be considered pristine infall galaxies that come from the same environment as the field. Galaxy pre-processing may still occur in groups that lie beyond this radius, but the large number of infalling intergroup galaxies will outweigh the influence of a minority of group galaxies beyond this radius.

\subsection{How dense environments affect star formation}
\subsubsection{Dense environments suppress star formation}
We have presented two pieces of evidence that suggest the densest environments in this protocluster suppress star formation: (i) the enhanced passive fraction within the groups relative to the intergroup region, and (ii) the global decrease in sSFR in the inner 3\,co-moving Mpc of the protocluster, where $\rho_{*}>10^{9}$\Msun\Mpc$^{-3}$ i.e. at least a factor of ten greater than the field.

This result is supported by previous work on this protocluster. For example, \citet{Quadri2012} found an excess of quiescent galaxies within a 1\,co-moving Mpc radius of the main group of this protocluster, whilst \citet{Tran2015} found the H$\alpha$-detected protocluster galaxies had suppressed SFRs within the central Mpc. In addition, several studies of clusters at a comparable  redshifts show similar results (e.g. \citealt{Cooke2016}, \citealt{Newman2014}). Our results show that the suppression of star formation extends beyond the nascent cluster core to the other massive groups, and so the suppression of the star formation is seen at large radii.

Although star formation is suppressed, the protocluster has a very high SFR density, especially in the most massive group of the protocluster. We have shown in Fig.\,\ref{fig:passive_fraction} that many protocluster galaxies are forming stars, with SFRs that correlate with their stellar mass. It therefore makes sense that large amounts of star formation accompany the excess of stellar mass that is the protocluster.

The high $\rho_{\rm SFR}$ is not an effect of the environment, but rather, this is the protocluster environment. The high $\rho_{*}$, and hence high $\rho_{\rm SFR}$, defines the visible baryonic component of the protocluster. This is the same situation as in the nearby Universe, where the $\rho_{\rm SFR}$ within clusters is typically two orders of magnitude greater than the field \citep{Stroe2015,Guglielmo2015} due to the high galaxy density.

 \subsubsection{Dense environments influence the mode of star formation}

In addition to suppressing star formation, the dense environment also enhances the relative fraction of red star-forming galaxies with respect to the blue star-forming population. The SED fits to the red star-forming galaxies suggest that the red colour is due to dust attenuation of the underlying light as they have high values of dust attenuation ($A_{2800}>2$\,mag). This is supported by the mid-infrared data as almost all of the red star-forming galaxies are detected in the {\it Spitzer}/MIPS image. Furthermore, the sSFRs of the star-forming galaxies in both environments are similar (KS probability = 0.1; Fig.\,\ref{fig:passive_fraction}) which suggests that the red star-forming galaxies are not simply passive galaxies that have been rejuvenated with some low-level recent star formation. 

This change in star formation mode, coupled with the high $\rho_{\rm SFR}$ in the protocluster, may be misinterpreted as an excess of star formation if only dusty star-forming galaxies are investigated. \clustername\ exhibits an excess of these red star-forming galaxies relative to the field by a factor of $\sim3$. This high fraction of dusty star-forming galaxies in the densest regions of the protocluster was discovered by \citet{Tran2010} and confirmed by \citet{Santos2014}. They interpreted these high fractions as a reversal in the SFR-density relation, from low SFR in high densities at low redshifts, to high SFRs in high densities at high redshifts. However, our results suggest a possible alternative explanation: that a larger fraction of the star forming galaxies contain large quantities of dust and are observed as obscured star forming galaxies. So the excess of dusty star forming galaxies is due to a change in the mode of star formation (from unobscured to obscured). The total fraction of star forming galaxies in the protocluster is not enhanced, but rather it has diminished.

\subsection{The lack of low-mass blue galaxies in groups}
\label{faint_blueies}
We observe a large number of blue galaxies with masses of $10^{9}-10^{9.5}$\Msun\ in the intergroup region that are not present in the massive groups (Fig.\,\ref{fig:CMassD}). These galaxies could be missing because they have red colours after being quenched or obscured. This would place them below the flux completeness limit of our $K-$selected catalogue, and hence they would not be detected.  This seems like the most natural explanation for the lack of low-mass blue galaxies as we find that the dense group environment quenches star formation, and enhances the fraction of obscured star forming galaxies. If all of these missing galaxies were red, the red fraction of $10^{9.3}$\Msun\ galaxies would be almost 80\%, whereas Fig.\,\ref{fig:red_fraction} shows that the red fraction of $10^{9.7}$\Msun\ galaxies is only $40-60$\% and the fraction decreases with decreasing stellar mass. Therefore, if these low-mass blue galaxies were quenched it would require more efficient quenching or a greater rate of obscuration occurring in these lower mass galaxies.

In addition, or alternatively, these galaxies may be stripped or have merged with the group galaxies.  \citet{Lotz2013} showed that minor mergers are common in \clustername\ with a merger rate that is $\sim3-10$ times higher than in the field. The missing stellar mass is less than 5\% of the stellar mass within the groups, so this mass could easily be hidden within a handful of the most massive group galaxies without significantly altering the high mass end of the galaxy stellar mass function.

\subsection{The evolution of environmental effects in protoclusters}

The fraction of protocluster galaxies that are satellites within a protocluster group increases with time \citep{Contini2016}. Since we only observe environmental effects within the dense groups of this protocluster, this suggests that environmental effects will become more prominent simply due to the hierarchical growth of clusters. However, this picture is too simplistic as the mechanism that quenches satellite galaxies may not be the same in the distant and local Universe \citep{Balogh2016}.

Beyond $z\sim2.5$ there are few massive groups within protoclusters \citep{Chiang2013}, so we expect the environmental effects we observe at $z=1.6$ to decrease at higher redshifts. The passive fraction of protoclusters at $z>2$ is difficult to measure, but there is some evidence that suggests that the suppression of star formation within protoclusters decreases at $z>2$ \citep{Cooke2014,Husband2016,Wang2016}. 

In contrast to this, the other environmental effect we measure does not seem to disappear at higher redshifts. The excess of red star-forming galaxies has been observed in protoclusters at even higher redshifts. \citet{Koyama2013} reported a population of red H$\alpha$ emitters that preferentially reside in the densest regions of the Spiderweb protocluster at $z=2.16$, whilst \citet{Hayashi2012} find the same in a $z=2.53$ protocluster. Thus the environmental process that influences this change of star formation mode is also at work at higher redshifts, when the groups are less massive. This suggests that separate processes are responsible for the suppression of star formation and the change of star formation mode in dense environments.

We end this discussion with some final words of caution. In this paper we have examined only a single protocluster at $z=1.6$. The large variation in cluster formation histories means that we should be wary of generalising the results from a single protocluster to the whole population. Furthermore, the environments within a protocluster vary with cluster growth and assembly. Therefore we should be particularly wary about  generalising these results to other redshifts.

\section{Conclusions}
\label{conclusions}

The goal of this work is to determine the effects of dense environments on galaxy evolution at high redshift. We examined the properties of galaxies in the \clustername\ protocluster at $z=1.6$, and compared them to the field. A third of the protocluster galaxies reside in dense groups, whilst two-thirds reside between the groups in an intermediate-density environment. 

We present evidence showing that:
\begin{itemize}
\item[(i)] the protocluster intergroup environment does not greatly influence the masses or colours of the galaxies. Intergroup protocluster galaxies appear to have the same properties as field galaxies. 
\item[(ii)] the protocluster groups have twice the red galaxy fraction of the intergroup region. This is due to an enhanced passive fraction (by a factor of 1.7) and a higher fraction of star forming galaxies exhibiting red colours (by a factor of 3). The enhancement of red and passive galaxies within groups is most prominent at low stellar masses, but the excess of red star-forming galaxies appears to be at the same level at all masses ($>10^{9.7}$\Msun).
\item[(iii)]  both the stellar mass density ($\rho_{*}$) and star formation rate density ($\rho_{\rm SFR}$) are greatly enhanced in the protocluster, particularly within the groups where the $\rho_{\rm SFR}$ is $\sim800$ times that of the field. Both $\rho_{*}$ and $\rho_{\rm SFR}$ decrease with distance from the protocluster core, until they reach the field level at $\sim6$\,co-moving Mpc. The total sSFR is suppressed in the central 3\,co-moving Mpc region compared to the outer regions due to the presence of several massive groups within this radius.
\end{itemize}
We conclude that galaxies are preprocessed within the groups of the protocluster. This dense environment suppresses star formation and enhances the fraction of star forming galaxies that are red.

\section{Acknowledgments}
We thank the anonymous referee for their comments which greatly improved this paper. NAH acknowledges support from STFC through an Ernest Rutherford Fellowship. EAC acknowledges support from STFC. SIM acknowledges the support of the STFC consolidated grant ST/K001000/1 to the astrophysics group at the University of Leicester. This work is based on observations made with ESO Telescopes at the La Silla Paranal Observatory under programme ID 089.A-0126.

\bibliographystyle{mn2e}\bibliography{References,mn-jour}
\label{lastpage}
\clearpage
\end{document}